\documentclass{article}

\usepackage{amssymb}
\usepackage[pdftex]{graphicx}

\begin{document}
\title{Ordered States in Fcc Kagome Antiferromagnets with Magnetic Dipolar Interactions}
\author{Terufumi Yokota\\
{\it College of Engineering, Nihon University}\\
{\it Koriyama, Fukushima, 963-8642, Japan}
}



\maketitle

\begin{abstract}
Ordered states for a classical Heisenberg model on fcc lattice structure with ABC stacked kagome planes of magnetic ions are investigated by numerically solving the Landau-Lifshitz (LL) equation. Both the nearest-neighbor exchange interaction and the magnetic dipolar interactions are included in the model. The model with only the nearest-neighbor antiferromagnetic exchange interaction is known to show the layered $120^{\circ}$ spin structure. On the other hand, the model with only the magnetic dipolar interactions is known to exhibit a continuous degeneracy expressed by six sublattice spin vectors, which is reduced by an order-by-disorder process with thermal fluctuations. In the present study, other ordered states appear for various values of relative strength of the two kinds of the interactions. A vortex spin structure on hexagonal lattice points in the kagomeplanes is a novel one. Another ordered state may be glassy state in which apparent translational symmetry cannot be seen. Layered $120^{\circ}$ spin structures but not uniform in the direction perpendicular to the kagome planes with various period in the direction appear depending on the relative strength of the two kinds of the interactions.    
\end{abstract}

\newpage
\section{Introduction}
In geometrically frustrated spin systems with interactions of antiferromagnetic nature among spins located on lattice points which constitute basic triangles, non-trivial spin structures can be realized. The fcc lattice structure with ABC stacked kagome planes of magnetic ions is an example of such frustrated systems. This lattice structure of magnetic ions is relevant to Mn in ${\rm Mn_3 Ir}$ alloy \cite{tetal} for example. The structure of stacked kagome planes was recognized in Ref. \cite{hetal}. In Fig. 1, the fcc kagome lattice structure with magnetic ions on cube faces forming ABC stacked kagome planes are depicted.
\begin{figure}
  \begin{center}
   \includegraphics[width=140mm]{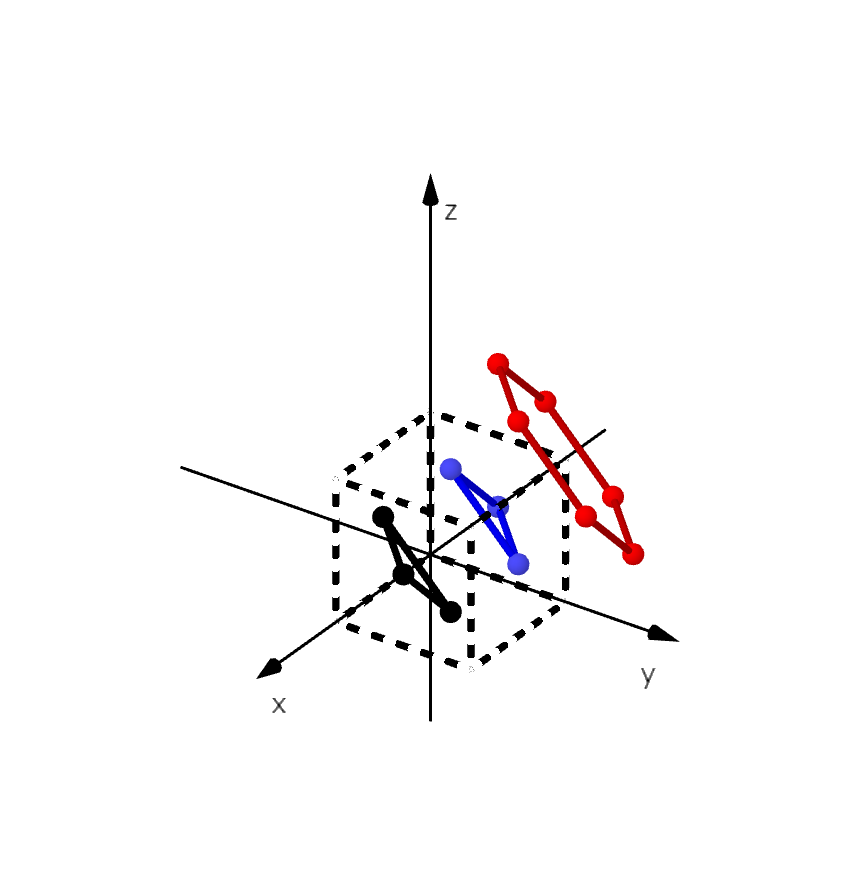}
  \end{center}
\vspace{-0.5cm}
\caption{The fcc kagome lattice structure with magnetic ions on cube faces forming ABC stacked kagome planes.}
\end{figure}
Nonmagnetic ions are at the corners. A part of the kagome lattice in a stacked plane is depicted in Fig. 2.
\begin{figure}
  \begin{center}
   \includegraphics[width=140mm]{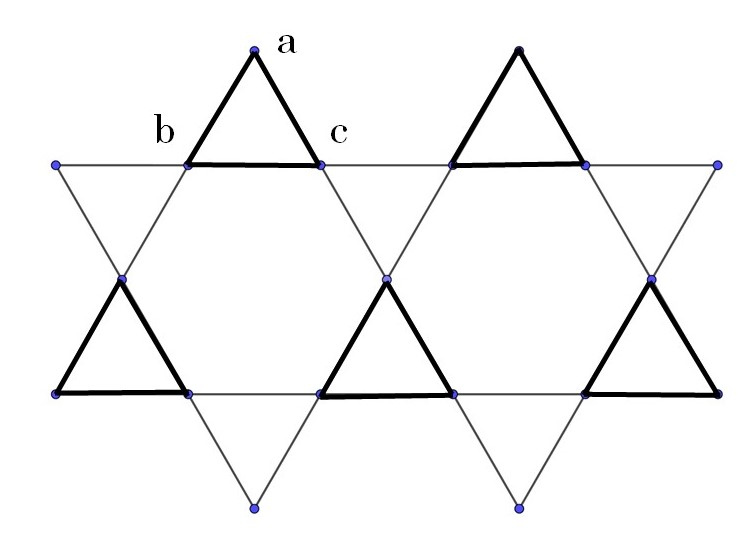}
  \end{center}
\vspace{-0.5cm}
\caption{A part of the kagome lattice in a stacked plane. All kagome lattice points can be divided into three sublattice points labeled by a, b, and c.}
\end{figure}
All kagome lattice points can be divided into three sublattice points belonging to the basic up-pointing triangles which is shown by the bold lines on the figure. Three sublattice points are labeled by a, b, and c in the figure for a triangle. Corresponding points for other triangles belong to a, b, and c sublattices, respectively.

In Ref.  \cite{hetal}, the Heisenberg model with nearest neighbor antiferromagnetic exchange interaction on the fcc kagome lattice was investigated by the Monte Carlo simulations. The ordered state is proposed to be a $120^{\circ}$ spin structure \cite{tetal} involving planes of defects. The spins are on planes perpendicular to a $<111>$ direction.

Ordered state of the Heisenberg model with only dipolar interactions on the fcc kagome lattice was studied in Ref. \cite{wetal} by analytic method and Monte Carlo simulations. There exists a continuous degeneracy which is reduced by thermal fluctuations at very low temperature with an order-by-disorder process. Spins on the fcc kagome lattice lie in a $\{111\}$ plane for the ordered state. Spins on alternate kagome planes belonging to the same sublattice change sign so that the order is period 2 in unit of distance between two consecutive kagome planes.

Because the ordered state for the system with only nearest neighbor antiferromagnetic interaction is rather different from that with only dipolar interactions, it would be an interesting problem to study the ground states for systems with both interactions. In this paper, ordered states for the Heisenberg model with the two kinds of interactions on the fcc kagome lattice are investigated by numerically solving the Landau-Lifshiths (LL) equation. Ordered states for only nearest neighbor antiferromagnetic exchange interaction and only dipolar interactions are also obtained and they are similar to what have been obtained in Refs. \cite{hetal} and \cite{wetal}, respectively although they are a bit different. Besides the two kinds of ordered states, might-be glassy states and states with longer period than 2 along $<111>$ directions are observed altering the relative value of the two kinds of interactions.

In Sec. 2, the antiferromagnetic Heisenberg model with dipole-dipole interactions on the fcc kagome lattice is introduced. The LL equation for the model is given. A continuous space approximation for the dipole-dipole interactions is used to make numerical calculations effectively. Ground states are obtained by numerically solving the LL equation for various values of the two kinds of interactions in Sec. 3. Discussion and conclusion are given in Sec. 4.

\section{Model and Equation}
Heisenberg antiferromagnet on the fcc kagome lattice with dipole-dipole interactions is described by the Hamiltonian
\begin{equation}
\mathcal{H}=\sum_{<i, j>}J_{ij}{\bf S}_i \cdot{\bf S}_j-\omega\sum_{<i, j>}\frac{3({\bf S}_i \cdot{\bf R}_{ij})({\bf S}_j \cdot{\bf R}_{ij})-{\bf S}_i \cdot{\bf S}_j R_{ij}^2}{R_{ij}^5}\equiv \mathcal{H}_{ex}+\mathcal{H}_{dip}
\end{equation}
where $i$ represents a lattice site of the fcc kagome lattice with the periodic boundary condition. The summation is taken over pairs of the fcc kagome lattice sites and $J_{ij}$ takes a non-negative value $J$ only for nearest-neighbor pairs. A cubic unit cell of the fcc kagome lattice is represented by broken lines in Fig. 1 where the size of the unit cell is  $2a\times 2a\times 2a$. In this paper, $a$ is set to 1. A classical three-dimensional spin with unit length ${\bf S}_i$ is located on the fcc kagome lattice points.
The first term of the Hamiltonian represents the nearest-neighbor antiferromagnetic interaction. The second term of the Hamiltonian represents long-range dipole-dipole interactions where ${\bf R}_{ij}$ is the vector from a site $i$ to a site $j$ and $\rm{R}_{ij} =\vert{\bf R}_{ij}\vert$.

The LL equation is
\begin{equation}
\frac{d{\bf S}_i (t)}{dt}=-{\bf S}_i (t)\times\left({\bf B}_i (t)+\gamma{\bf S}_i (t)\times{\bf B}_i (t)\right)
\end{equation}
where ${\bf B}_i (t)=-\frac{\delta\mathcal{H}}{\delta{\bf S}_i (t)}+{\bf \xi }_i (t)$ and $\gamma$ is the damping constant. ${\bf \xi }_i$ is Gaussian thermal noise, which satisfies
\begin{eqnarray}
&&<{\bf \xi }_i (t)>=0\nonumber\\
&&<\xi^a_i (t)\xi^b_j (t')>=\delta_{ab}\delta_{ij}\delta (t-t')\frac{2\gamma}{1+\gamma^2}T
\end{eqnarray}
where $<\cdots >$ is the average, $T$ is the temperature, and the index $a$ represents the vector component \cite{gl}. $\delta_{ab}$ and $\delta_{ij}$ are the Kronecker deltas and $\delta (t-t')$ is the Dirac delta function.

Fourier transform of the LL equation will be used to properly include the long-range dipole-dipole interactions in the system with periodic boundary condition. A continuous space approximation is used to obtain an explicit form of the dipole-dipole interaction part in the equation.

The exchange part of the equation is
\begin{equation}
{\bf B}_{i, ex} (t)=-\frac{\delta\mathcal{H}_{ex}}{\delta{\bf S}_i (t)}=-\sum_{j (n. n. {\rm of} i)}J{\bf S}_j
\end{equation}
where the summation is taken over the nearest neighbor sites of $i$. The discrete Fourier transform of the exchange part for the lattice size of $L\times L\times L$ is 
\begin{equation}
{\bf B}_{ex} ({\bf K})=-\frac{8}{3}J{\bf S}({\bf K})(\cos K_x \cos K_y+\cos K_x \cos K_z+\cos K_y \cos K_z )
\end{equation} 
where ${\bf S}({\bf K})$ is the Fourier transform of ${\bf S}_i$ and $K_A =\frac{2\pi m_A}{L}$. $m_A$ is an integer that satisfies $-\frac{L}{2}<m_A \le \frac{L}{2}$. The form of ${\bf B}_{ex} ({\bf K})$ is the same as those for the fcc and the pyrochlore lattices. Only the coefficient is different, that is $-4$ and $-2$ for the fcc lattice \cite{y1} and the pyrochlore lattice \cite{y2}, respectively. The coefficients are proportional to the ratio of magnetic ions in the fcc lattice points, which is the base structure of the fcc kagome and the pyrochlore lattices.

The dipole-dipole term for the equations is
\begin{equation}
{\bf B}_{i, dip}=-\frac{\delta\mathcal{H}_{dip}}{\delta{\bf S}_i }.
\end{equation}
The Fourier transform is obtained using the continuous space approximation. The three components of ${\bf B}_{i, dip}$ in the continuous space approximation are given by
\begin{equation}
\label{Bx}
B^x_{i, dip}=\omega\int d{\bf R}'[S^x({\bf R}')G_x({\bf R}-{\bf R}')+3S^y({\bf R}')G_{xy}({\bf R}-{\bf R}')+3S^z({\bf R}')G_{xz}({\bf R}-{\bf R}')],\\
\end{equation} 
\begin{equation}
\label{By}
B^y_{i, dip}=\omega\int d{\bf R}'[S^y({\bf R}')G_y({\bf R}-{\bf R}')+3S^x({\bf R}')G_{xy}({\bf R}-{\bf R}')+3S^z({\bf R}')G_{yz}({\bf R}-{\bf R}')],\\
\end{equation} 
and
\begin{equation}
\label{Bx}
B^z_{i, dip}=\omega\int d{\bf R}'[S^z({\bf R}')G_z({\bf R}-{\bf R}')+3S^x({\bf R}')G_{xz}({\bf R}-{\bf R}')+3S^y({\bf R}')G_{yz}({\bf R}-{\bf R}')]\\
\end{equation} 
where $G_x ({\bf R})=\frac{3x^2-R^2}{R^5}$, $G_y ({\bf R})=\frac{3y^2-R^2}{R^5}$, $G_z ({\bf R})=\frac{3z^2-R^2}{R^5}$, $G_{xy} ({\bf R})=\frac{xy}{R^5}$, $G_{xz} ({\bf R})=\frac{xz}{R^5}$, and $G_{yz} ({\bf R})=\frac{yz}{R^5}$. ${\bf R}=(x, y, z)$ and $R=\vert{\bf R}\vert$.  Discrete Fourier transforms with the form of the continuous space approximation are used to include the long-range dipole-dipole interaction. Explicit forms will be given shortly. The Fourier transforms on the fcc kagome lattice are given by
\begin{eqnarray}
B^x_{dip}({\bf K})&=&\omega\sum_{m'_x , m'_y, m'_z}\epsilon ({\bf K}-{\bf K}')\left[S^x ({\bf K}')G_x ({\bf K}')+3S^y ({\bf K}')G_{xy}({\bf K}')\right.\nonumber\\
&&\hspace{1cm}
\left.+3S^z ({\bf K}')G_{xz}({\bf K}')\right],
\end{eqnarray} 
\begin{eqnarray}
B^y_{dip}({\bf K})&=&\omega\sum_{m'_x , m'_y, m'_z}\epsilon ({\bf K}-{\bf K}')\left[S^y ({\bf K}')G_y ({\bf K}')+3S^x ({\bf K}')G_{xy}({\bf K}')\right.\nonumber\\
&&\hspace{1cm}
\left.+3S^z ({\bf K}')G_{yz}({\bf K}')\right],
\end{eqnarray} 
and
\begin{eqnarray}
B^z_{dip}({\bf K})&=&\omega\sum_{m'_x , m'_y, m'_z}\epsilon ({\bf K}-{\bf K}')\left[S^z ({\bf K}')G_z ({\bf K}')+3S^x ({\bf K}')G_{xz}({\bf K}')\right.\nonumber\\
&&\hspace{1cm}
\left.+3S^y ({\bf K}')G_{yz}({\bf K}')\right],
\end{eqnarray}
where $G({\bf K})$'s are the Fourier transforms of  $G({\bf R})$'s and $\epsilon ({\bf K})$ is
\begin{eqnarray}
\epsilon ({\bf K})&=&\frac{1}{8}\left[3{\bf\delta}_{{\bf K}, {\bf 0}}+3{\bf\delta}_{{\bf K}, {\bf\Pi}}-{\bf\delta}_{{\bf K}, (\pi, 0, 0)}-{\bf\delta}_{{\bf K}, (0, \pi, 0)}-{\bf\delta}_{{\bf K}, (0, 0, \pi)}-{\bf\delta}_{{\bf K}, (0, \pi, \pi)}\right.\nonumber\\
&&\left.-{\bf\delta}_{{\bf K}, (\pi, 0, \pi)}-{\bf\delta}_{{\bf K}, (\pi, \pi, 0)}\right]
\end{eqnarray}
where ${\bf 0}=(0, 0, 0)$ and ${\bf\Pi}=(\pi, \pi, \pi)$. Product of three Kronecker deltas is represented by ${\bf\delta}$ and ${\bf\delta}_{{\bf K}, {\bf 0}}=\delta_{k_x , 0}\delta_{k_y , 0}\delta_{k_z , 0}$ for example. $G({\bf K})$'s are obtained explicitly performing the integral in the polar coordinate system and given by
\begin{equation}
G_x({\bf K})=4\pi \left[1-3\left(\frac{K_x}{K}\right)^2 \right]\frac{\sin (Kd)-Kd\cos (Kd)}{(Kd)^3},
\end{equation} 
\begin{equation}
G_y({\bf K})=4\pi \left[1-3\left(\frac{K_y}{K}\right)^2 \right]\frac{\sin (Kd)-Kd\cos (Kd)}{(Kd)^3},
\end{equation} 
\begin{equation}
G_z({\bf K})=4\pi \left[1-3\left(\frac{K_z}{K}\right)^2 \right]\frac{\sin (Kd)-Kd\cos (Kd)}{(Kd)^3},
\end{equation} 
\begin{equation}
G_{xy}({\bf K})=-4\pi \frac{K_x K_y}{K^2}\cdot\frac{\sin (Kd)-Kd\cos (Kd)}{(Kd)^3},
\end{equation} 
\begin{equation}
G_{xz}({\bf K})=-4\pi \frac{K_x K_z}{K^2}\cdot\frac{\sin (Kd)-Kd\cos (Kd)}{(Kd)^3},
\end{equation} 
and
\begin{equation}
G_{yz}({\bf K})=-4\pi \frac{K_y K_z}{K^2}\cdot\frac{\sin (Kd)-Kd\cos (Kd)}{(Kd)^3}
\end{equation} 
where $K=\vert {\bf K}\vert$ and $d$ is the cutoff length that is the lower limit for the dipole-dipole interactions. In this paper, $d$ is set to $1$. 

\section{Various Ordered States}
The LL equation is numerically solved to obtain ground states for different values of $\omega /J$. Implicit Gauss-Seidel  method with fractional step \cite{wge} is used to improve numerical stability. The system size is $64\times 64\times 64$ with the periodic boundary condition that approximates the infinite system. Discrete Fourier transforms with the forms of the continuous space approximation for the dipole-dipole interaction part in Eqs. (14) - (19) are used. The damping constant is set as $\gamma =1.0$. The initial directions of the spins are randomly chosen, which corresponds to a paramagnetic state at high temperature. The temperature is set to $25.0$ or $100.0$ at the time $t=0.0$ not so as to form a pattern at the initial temperature. The temperature is lowered at a speed of $dT^{1/2} /dt =-10^{-4}$ to $T=0.0$ at $t=50,000$ or $100,000$. Spin structures ${\bf S}({\bf R})$ are obtained. Types of order are classified by investigating distributions of angles among spin directions in the basic up-pointing triangles depicted in Fig. 2 and other quantities.

Besides the $120^{\circ}$ spin structure seen in the system with only the nearest neighbor antiferromagnetic interaction \cite{hetal} and the ordered state with the opposite spin directions on alternate kagome planes along the $<111>$ directions in the system with only dipolar interactions, several kinds of ordered states appear in systems with both kinds of interactions depending on the ratio of the two interactions.

\subsection{$120^{\circ}$ spin structure}
$120^{\circ}$ spin structures \cite{hetal} are observed for $0.0\le\omega /J \le 1.0$ although some variations exist as shown below.

For $\omega /J =0.0$ with $\omega =0.0$ and $J=2.0$, $120^{\circ}$ planar order with defect lines (planes) appears. In Fig. 3, spin structures of a kagome plane in the central part of the fcc kagome lattice are shown. 
\begin{figure}
 \begin{minipage}{0.10\hsize}
  \begin{center}
   \includegraphics[width=14mm]{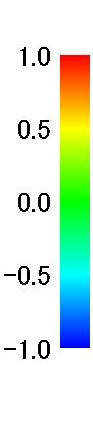}
  \end{center}
 \end{minipage}
 \begin{minipage}{0.44\hsize}
  \begin{center}
   \includegraphics[width=55mm]{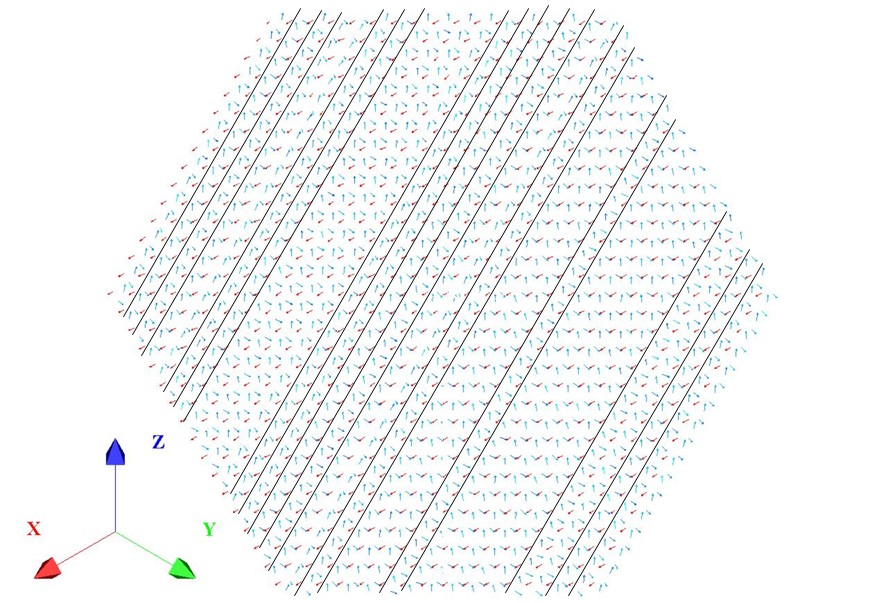}
  \end{center}
 \end{minipage}
 \begin{minipage}{0.44\hsize}
  \begin{center}
   \includegraphics[width=55mm]{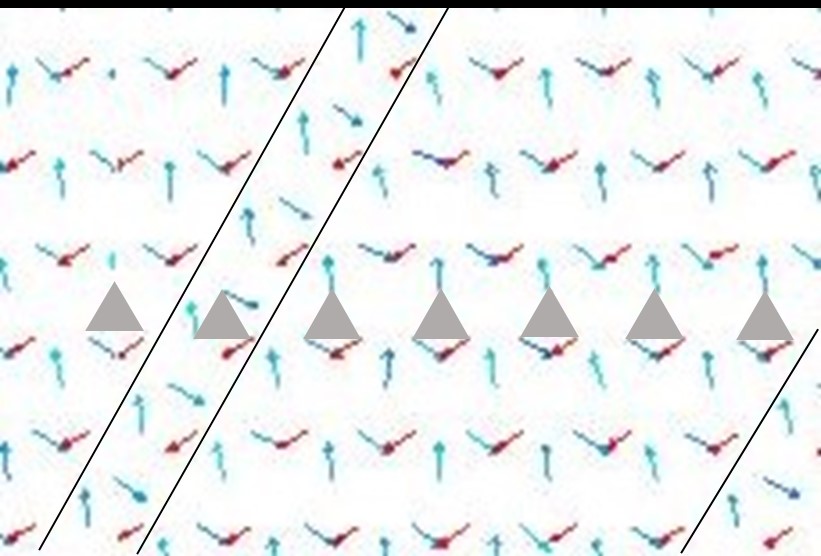}
  \end{center}
 \end{minipage}
\vspace{1.0cm}
\caption{Spin structure of a kagome plane for  $\omega =0.0$ and $J=2.0$ (left). Lines are defect lines. The color legend shows the value of $S^x$. A part of the left figure is enlarged and depicted in the right figure. A row of up-pointing triangles of which spins are at vertexes is shown.}
\end{figure}
The color legend shows the value of $S^x$. Defect lines are also depicted. There are 18 defect lines in the figure. A part of the left figure is enlarged, in which a row of up-pointing triangles is depicted. Spins exist at vertexes of the triangles. Two spin directions of a triangle between the two defect lines are different from those at the opposite sides of the defect lines. Angles between spin directions in each triangle  are $120^{\circ}$. To distinguish the spin directions of up-pointing triangles, it would be convenient to introduce the vector "Chirality" \cite{ms} ${\bf\kappa}$ defined by
\begin{equation}
{\bf \kappa}=\frac{2}{3\sqrt 3}\left({\bf S}^a \times {\bf S}^b +{\bf S}^b \times {\bf S}^c +{\bf S}^c \times {\bf S}^a \right).
\end{equation}
The direction of ${\bf \kappa}$ in a side of a defect line is opposite to that in another side of the line. The "chirality" vectors ${\bf \kappa}$ on the kagome plane depicted in Fig. 3 are ${\bf \kappa}_1 \simeq (0.00, -0.63, 0.77)$ and ${\bf \kappa}_2 \simeq (0.00, 0.63, -0.77)$, and ${\bf \kappa}_1 \simeq -{\bf \kappa}_2$ These chirality vectors are not along the $<111>$ directions, which results from the fact that the spins ${\bf S}^a$, ${\bf S}^b$, and ${\bf S}^c$ are not on the $\{111\}$ kagome planes although the spins are approximately on a plane. These qualitative features are the same for the $L=32$ system. This is in contrast to the spin directions in the systems with dipolar interactions as explained below.

As the dipolar interactions are switched on, defect lines (planes) disappear in the system size investigated and the chirality vectors are aligned along the $<111>$ direction, which is caused by the spins being on the $\{111\}$ kagome plane. The spin structure for $J=2.0$ and $\omega =0.01$ is shown in Fig. 4.
\begin{figure}
  \begin{center}
   \includegraphics[width=140mm]{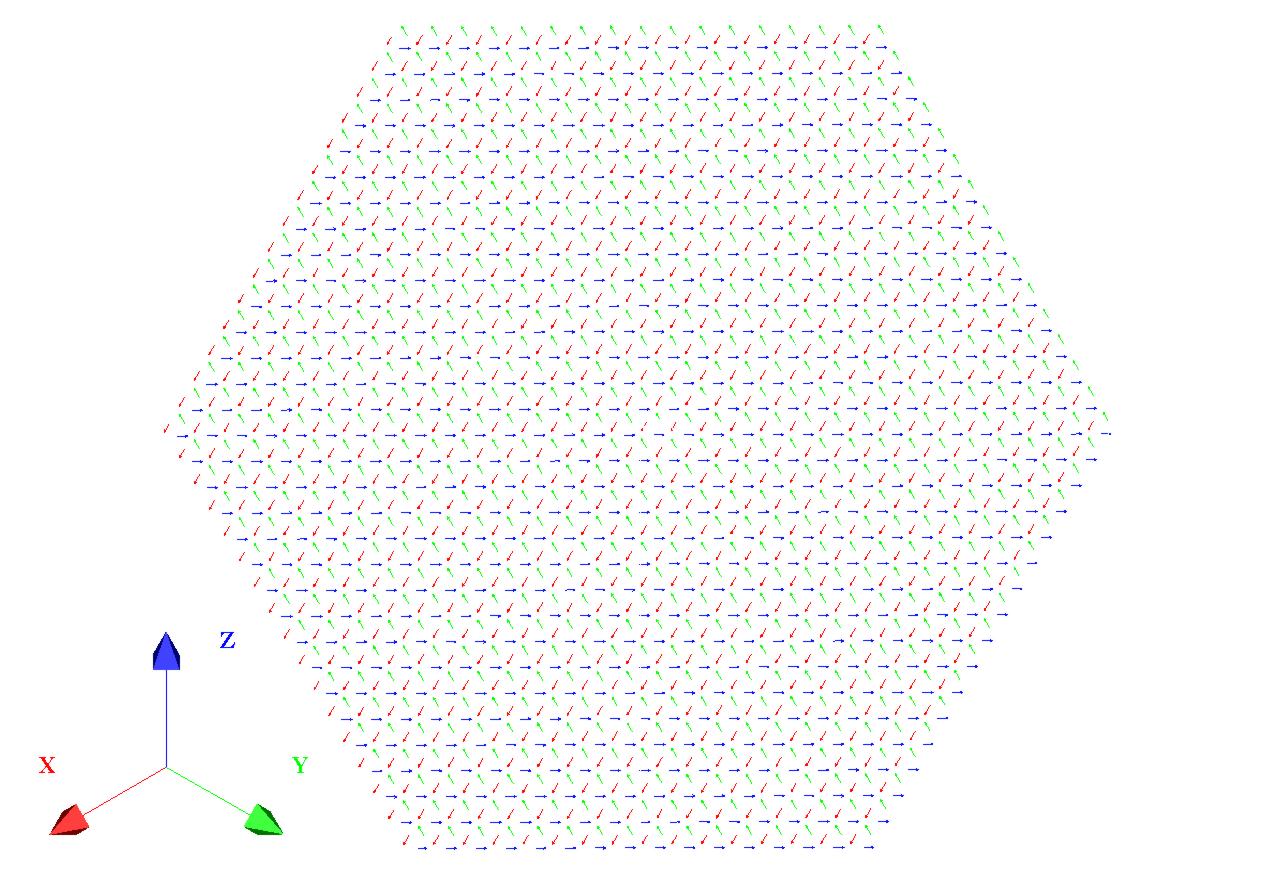}
  \end{center}
\vspace{-0.5cm}
\caption{Spin structure for $J=2.0$ and $\omega =0.01$. The colors represent the values of $S^x$ in the same way as Fig. 3. }
\end{figure}
The colors represent the values of $S^x$ in the same way as Fig. 3. The spins keep the  $120^{\circ}$ structure on the kagome planes. The same spin structure can be seen for other five systems with larger values of $\omega /J$, namely $0.05\le\omega /J \le 1.0$.

For the system between the pure exchange interaction and $\omega /J =0.005$, that is  $\omega /J =0.0005$ with $\omega =0.001$ and $J=2.0$, the spins are in a $120^{\circ}$ structure but the spin structure is different. There are 10 defect lines in the central kagome plane and spins with one $\kappa$ is different to that with another $\kappa$. Namely two $\kappa$'s for $L=64$ and $L=32$ form angles about $118^{\circ}$ and $159^{\circ}$, respectively and $\kappa$'s do not seem to be related to the crystal axes. 

\subsection{Spin structure determined in two triangles on successive kagome planes}
In the system with only dipolar interactions, ordered state by an order-by-disorder process is realized \cite{wetal}. The spin structure of the fcc kagome lattice is fully characterized by the six spin vectors on two up-pointing triangles in two successive kagome planes. The spin vectors on one triangle are \cite{wetal}
\begin{eqnarray}
{\bf S}_a &=&(\sin\theta\cos\phi, \sin\theta\sin\phi, \cos\theta),\nonumber\\
{\bf S}_b &=&(\alpha, \beta, -S_a^x),\nonumber
\end{eqnarray}
and
\begin{equation}
\label{magDip}
{\bf S}_c =(-\beta, -S_a^z -\alpha,  -S_a^y)
\end{equation}
where $\alpha =[2(S_a^x )^2 -1]/(2S_a^z)$ and $\beta =\pm\sqrt{1-(S_a^x )^2 -\alpha^2}$. The spin vectors on another triangle above or below on an adjacent $(111)$ kagome plane are 
\begin{equation}
\label{magDip2}
{\bf S}'_a = -{\bf S}_a,
\hspace{0.5cm}
{\bf S}'_b = -{\bf S}_b,
\hspace{0.5cm}
{\bf S}'_c = -{\bf S}_c.
\end{equation}

In the ordered state obtained by an order-by-disorder process, $\theta =\pi/2$ and $\phi =(2n+1)/4\pi$ where $n$ is an integer \cite{wetal}. The numerically obtained spin vectors of the triangles in a $(111)$ kagome plane of the central part of the fcc kagome lattice for $J=0.0$ and $\omega =0.5$ are
\begin{eqnarray}
{\bf S}_a \simeq (-0.710\pm 0.001, 0.702\pm 0.001, -0.039\pm 0.029),\nonumber\\
{\bf S}_b \simeq (-0.131\pm 0.044, -0.690\pm 0.009, 0.710\pm 0.001),\nonumber
\end{eqnarray}
and
\begin{equation}
\label{magDipR}
{\bf S}_c \simeq (0.690\pm 0.008, 0.170\pm 0.030, -0.702\pm 0.001).
\end{equation}
 These vectors are similar to  (\ref{magDip}) with $\theta =\pi /2$ and $\phi =3\pi /4$ although some deviations exist as $\theta =92.4^{\circ}$ and $\phi =135.3^{\circ}$. The spin structure represented by  (\ref{magDip}) with $\theta =\pi /2$ and $\phi =(2n+1)\pi /4$ is the $120^{\circ}$ structure. The three angles between the three spin vectors obtained from(\ref{magDipR}) are $\theta_{ab}\simeq 115^{\circ}$, $\theta_{bc}\simeq 135^{\circ}$, and $\theta_{ca}\simeq 110^{\circ}$. The deviation of the numerical result may be caused by the smallness of the system used in the calculations. In the smaller system with $L=32$,  $\theta =87.1^{\circ}$ and $\phi =313.5^{\circ}$ and the deviation from  $\theta =\pi/2$ and $\phi =(2n+1)/4\pi$ is a little larger than that for $L=64$.

This kind of the ordered state survives by inclusion of small nearest neighbor antiferromagnetic exchange interaction. The systems for $J=0.01, 0.02$, and $0.05$ with $\omega =0.5$ show similar spin vectors to those represented by the angles  $\theta =\pi/2$ and $\phi =(2n+1)/4\pi$. The deviation from (\ref{magDip}) with $\theta =\pi/2$ and $\phi =(2n+1)/4\pi$ become larger as the exchange interaction $J$ increases. For $J=0.05$, the three angles between the three vectors corresponding to (\ref{magDipR}) are $\theta_{ab}\simeq 113^{\circ}$, $\theta_{bc}\simeq 138^{\circ}$, and $\theta_{ca}\simeq 108^{\circ}$ and deviate more from  $120^{\circ}$ than those for smaller $J$'s.

\subsection{Glassy spin cluster or small multidomain state}
For larger nearest neighbor antiferromagnetic exchange interactions, the ordered states do not seem to order regularly in space. Spin structure on the central kagome plane for $J=0.1$ and $\omega =0.5$ is depicted in Fig. 5. The colors represent the values of $S^x$ in the same way as Fig. 3.     
\begin{figure}
  \begin{center}
   \includegraphics[width=140mm]{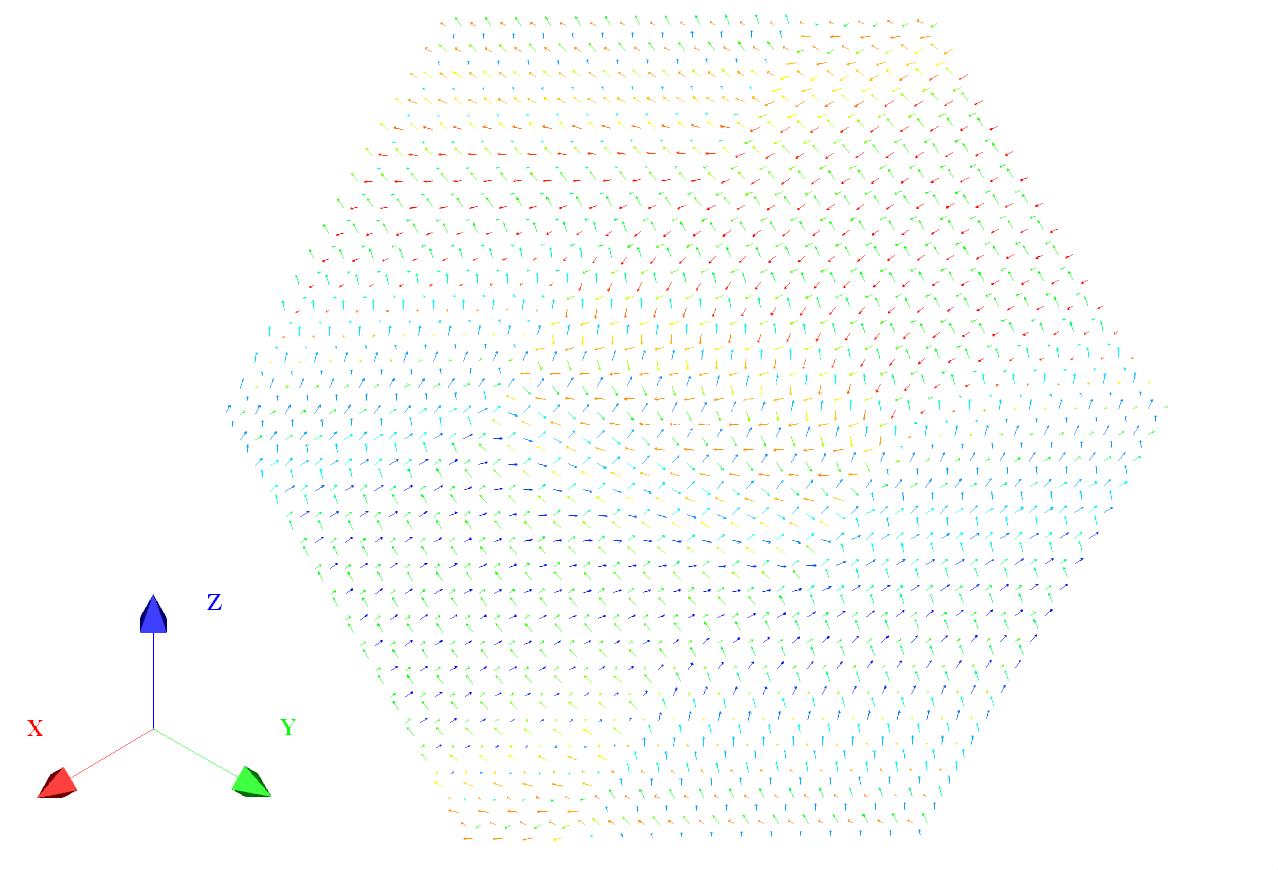}
  \end{center}
\vspace{-0.5cm}
\caption{Spin structure for $J=0.1$ and $\omega =0.5$. The colors represent the values of $S^x$ in the same way as Fig. 3. }
\end{figure}
The distributions of three angles between the three spin vectors of up-pointing triangles $\theta_{ab}, \theta_{bc}$, and $\theta_{ca}$ are very different from those in the previous subsection.

In Fig. 6, the distributions for the central $(111)$ and $(\overline{1}11)$ planes and those of $J=0.0$ and $\omega =0.5$ are depicted from left to right.  
\begin{figure}
 \begin{minipage}{0.24\hsize}
  \begin{center}
   \includegraphics[width=30mm]{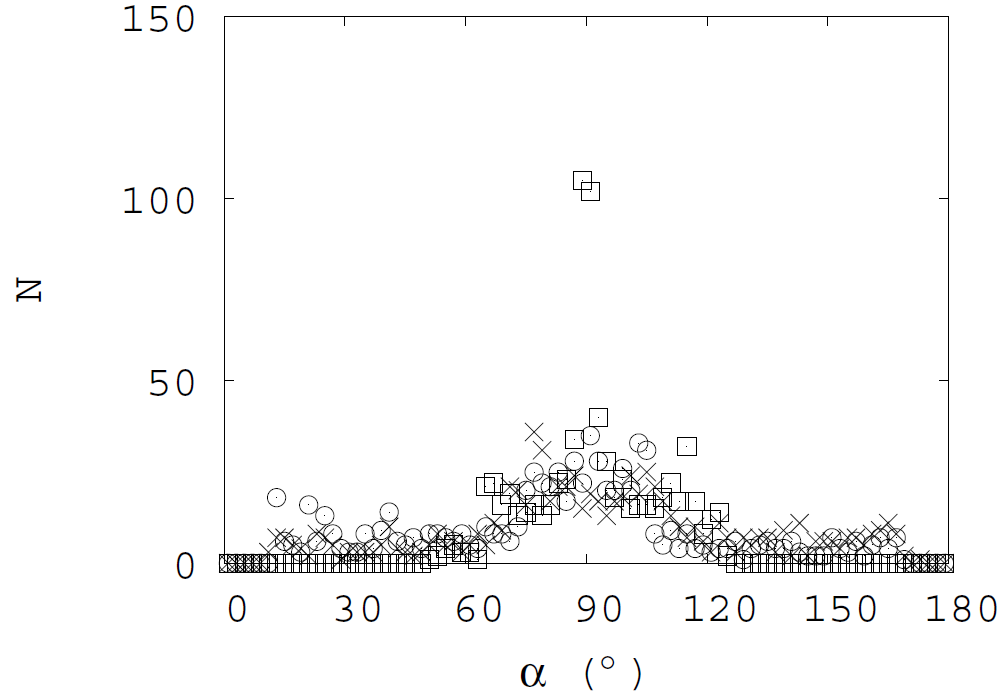}
  \end{center}
 \end{minipage}
 \begin{minipage}{0.24\hsize}
  \begin{center}
   \includegraphics[width=30mm]{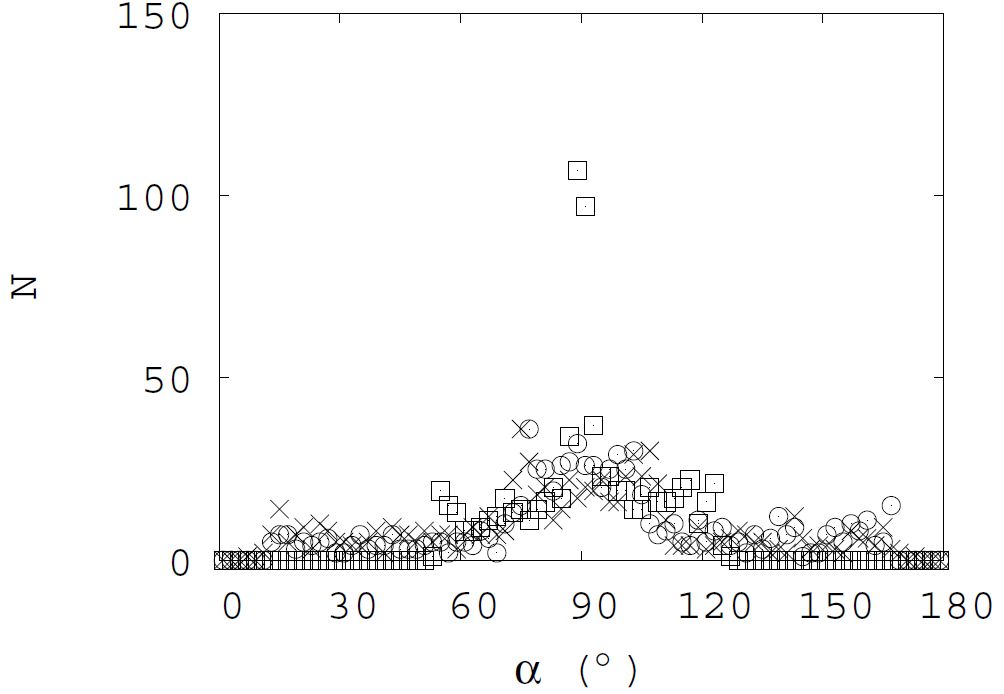}
  \end{center}
 \end{minipage}
 \begin{minipage}{0.24\hsize}
  \begin{center}
   \includegraphics[width=30mm]{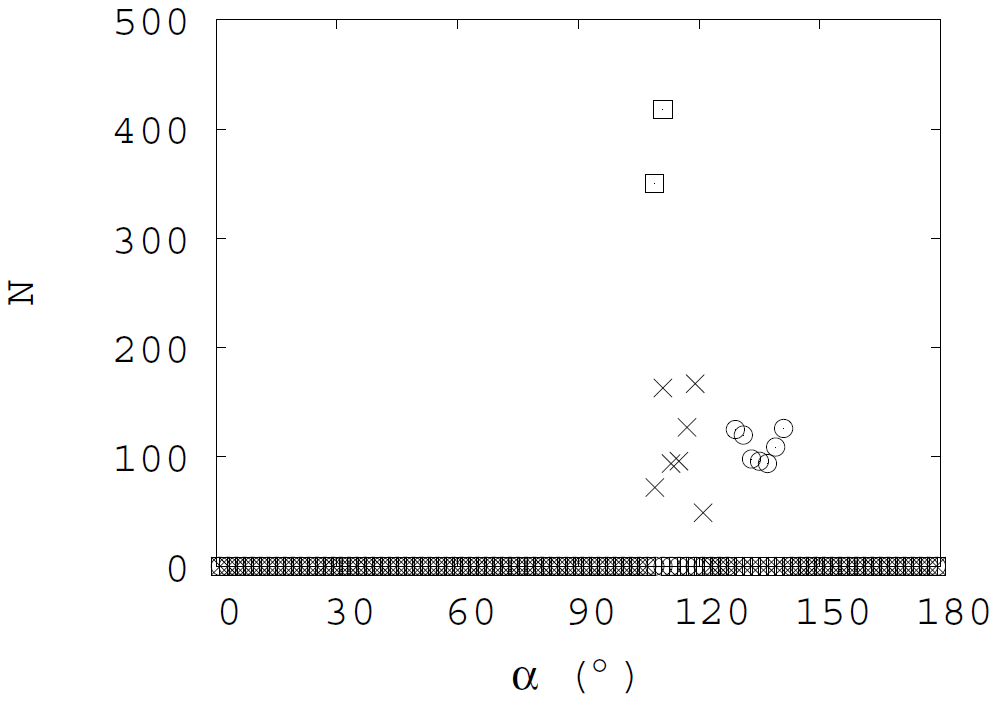}
  \end{center}
 \end{minipage}
 \begin{minipage}{0.24\hsize}
  \begin{center}
   \includegraphics[width=30mm]{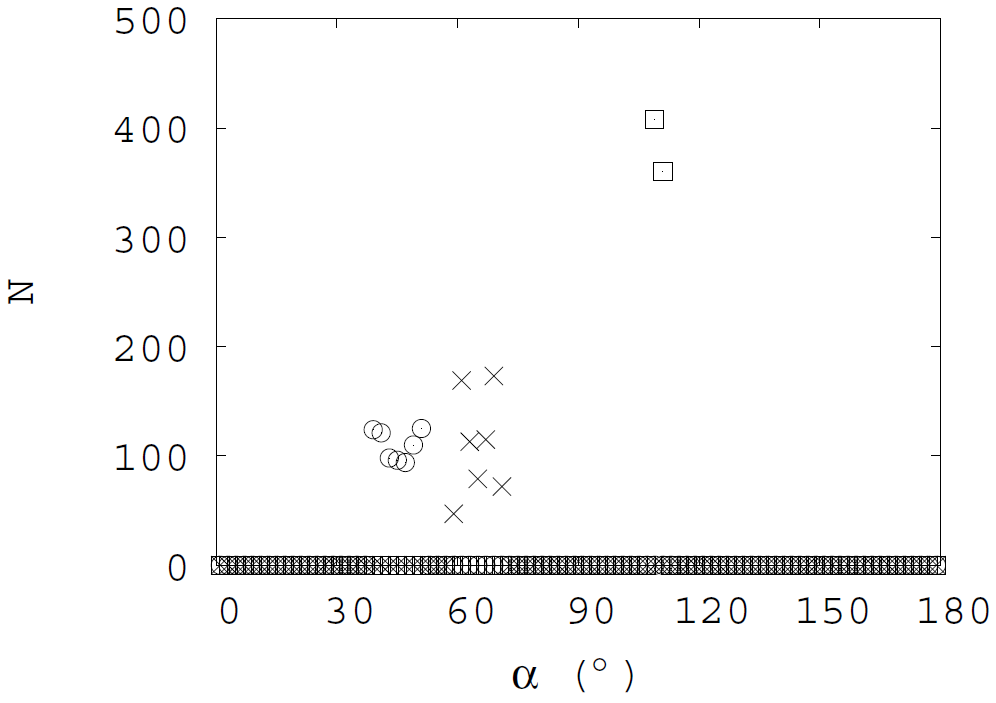}
  \end{center}
 \end{minipage}
\vspace{1.0cm}
\caption{Distributions of three angles between the three spin vectors of up-pointing triangles $\theta_{ab}, \theta_{bc}$, and $\theta_{ca}$for the central $(111)$ and $(\overline{1}11)$ planes for $J=0.1$ and $\omega =0.5$ and those for $J=0.0$ and $\omega =0.5$ are depicted from left to right. Crosses, circles, and squares represent angles for $\theta_{ab}$; $\theta_{bc}$, and $\theta_{ca}$, respectively in $2^\circ$ increments.}
\end{figure}
Crosses, circles, and squares represent angles for $\theta_{ab}, \theta_{bc}$, and $\theta_{ca}$, respectively in  $2^{\circ}$ increments. The distributions for the $(111)$ plane of $J=0.0$ and $\omega =0.5$ have peaks around $120^{\circ}$ although some deviations exist as stated in the previous subsection. The distributions for the $(\overline{1}11)$ plane have peaks around $120^{\circ}$ and $60^{\circ}$. The distributions for $J=0.1$ and $\omega =0.5$ are very different from those for the pure dipolar system and they have broad peaks around $90^{\circ}$ both for the $(111)$ and $(\overline{1}11)$ planes.

The spin vectors in the previous subsection are staggered as expressed in (\ref{magDip}) and  (\ref{magDip2}). This feature seems to be kept partially for $J=0.1$ and $\omega =0.5$ although distributions of some spin components are too broad to see the feature. In Fig. 7, the distributions of $S_a^y$ for the successive four kagome layers are depicted in $0.01$ increments.
\begin{figure}
 \begin{minipage}{0.24\hsize}
  \begin{center}
   \includegraphics[width=30mm]{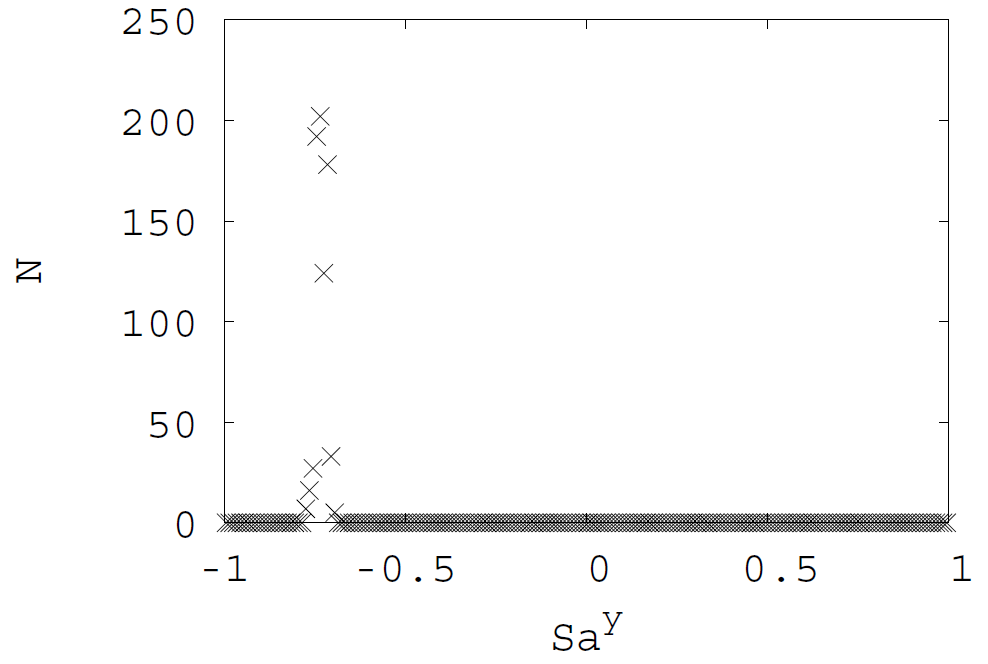}
  \end{center}
 \end{minipage}
 \begin{minipage}{0.24\hsize}
  \begin{center}
   \includegraphics[width=30mm]{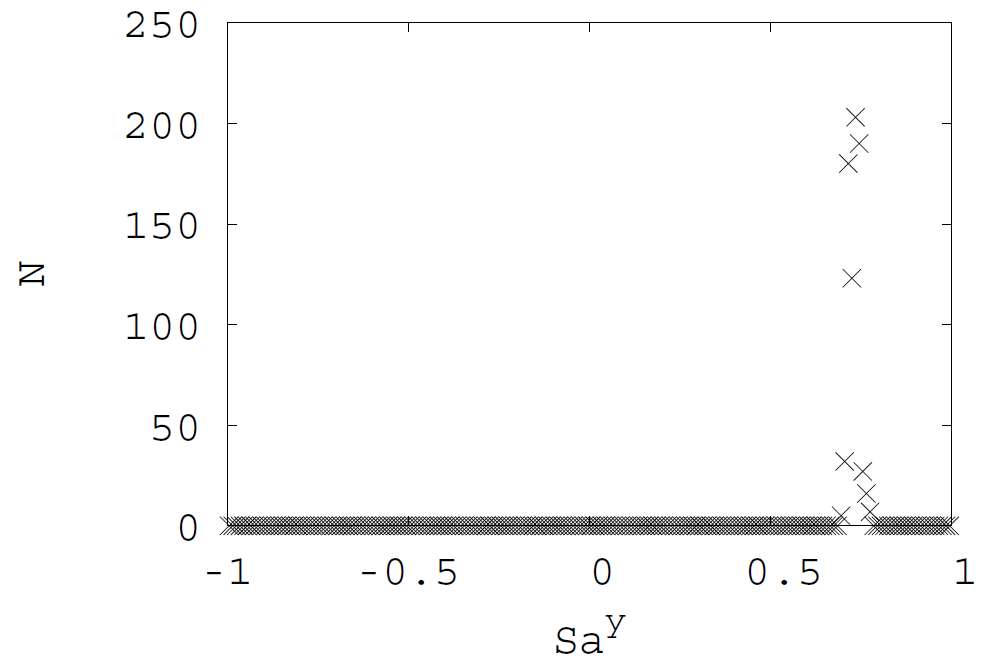}
  \end{center}
 \end{minipage}
 \begin{minipage}{0.24\hsize}
  \begin{center}
   \includegraphics[width=30mm]{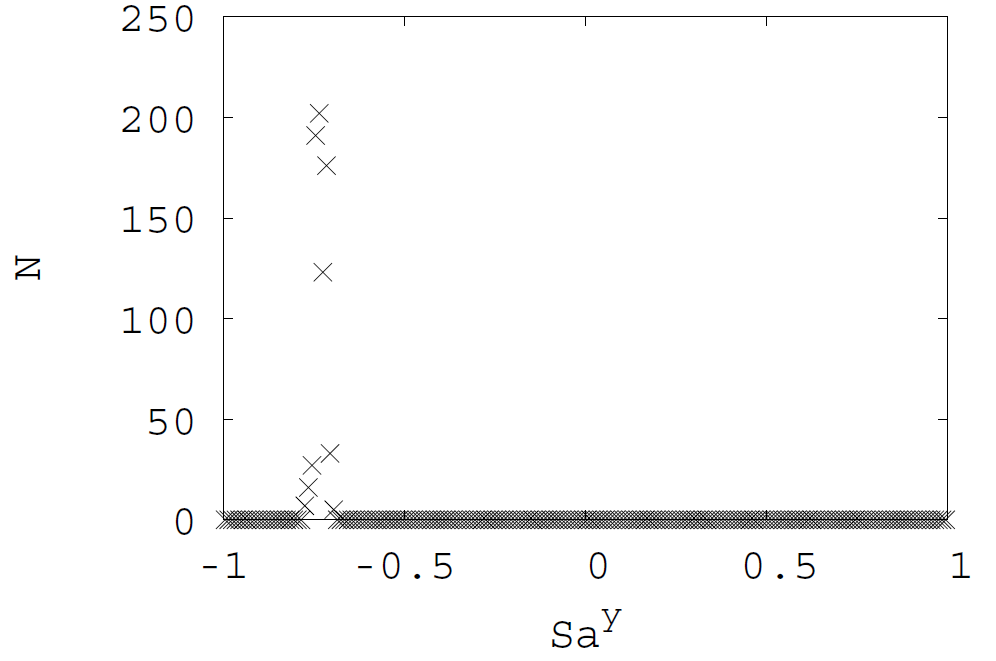}
  \end{center}
 \end{minipage}
 \begin{minipage}{0.24\hsize}
  \begin{center}
   \includegraphics[width=30mm]{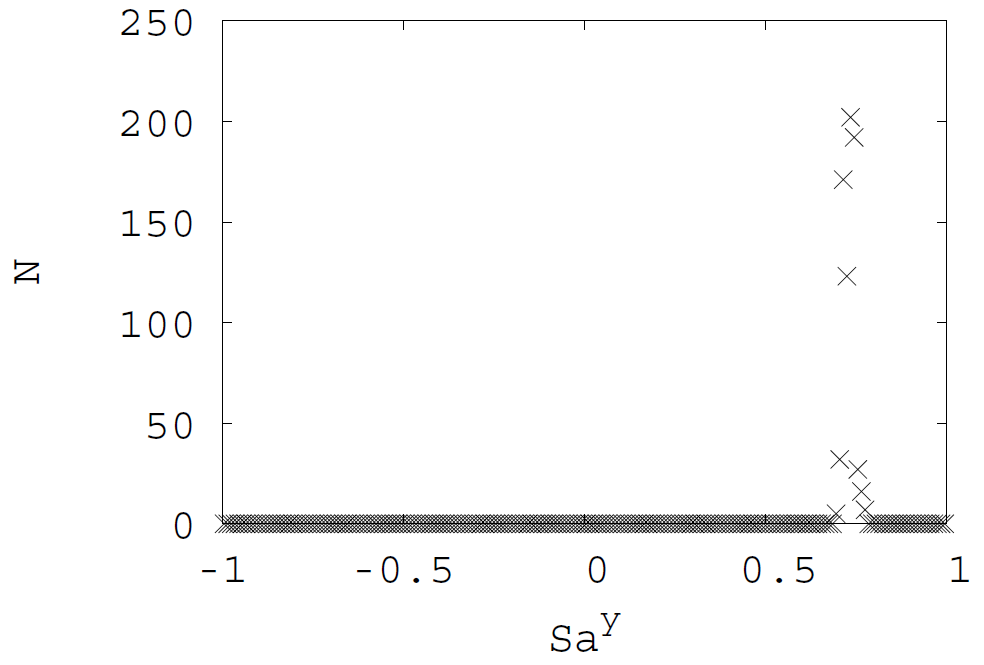}
  \end{center}
 \end{minipage}
\vspace{1.0cm}
\caption{Distributions of $S_a^y$ of the successive four kagome layers for $J=0.1$ and $\omega =0.5$.}
\end{figure}
These distributions show rather sharp peaks and the $S_a^y$ peak values are staggered for the kagome planes. Because the distributions for other spin components are broad, the spin vectors are not decided by the set of six spin values as  (\ref{magDip}) and  (\ref{magDip2}). However, also in this case, the staggered nature of spin values seems to remain as the distributions of $S_a^x$ for two successive kagome layers depicted in Fig. 8 show.
\begin{figure}
 \begin{minipage}{0.49\hsize}
  \begin{center}
   \includegraphics[width=65mm]{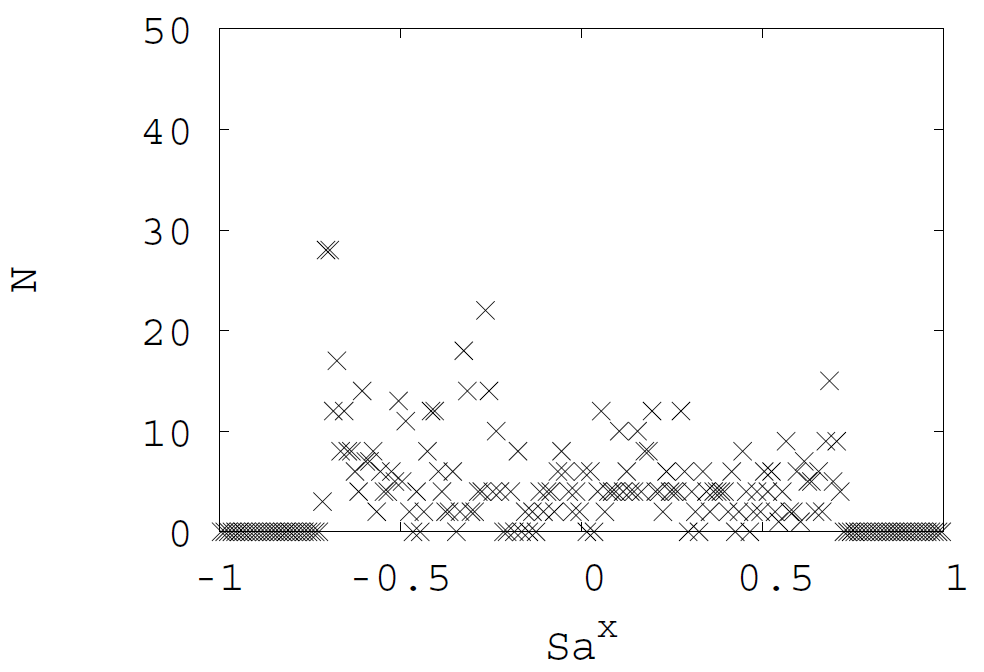}
  \end{center}
 \end{minipage}
 \begin{minipage}{0.49\hsize}
  \begin{center}
   \includegraphics[width=65mm]{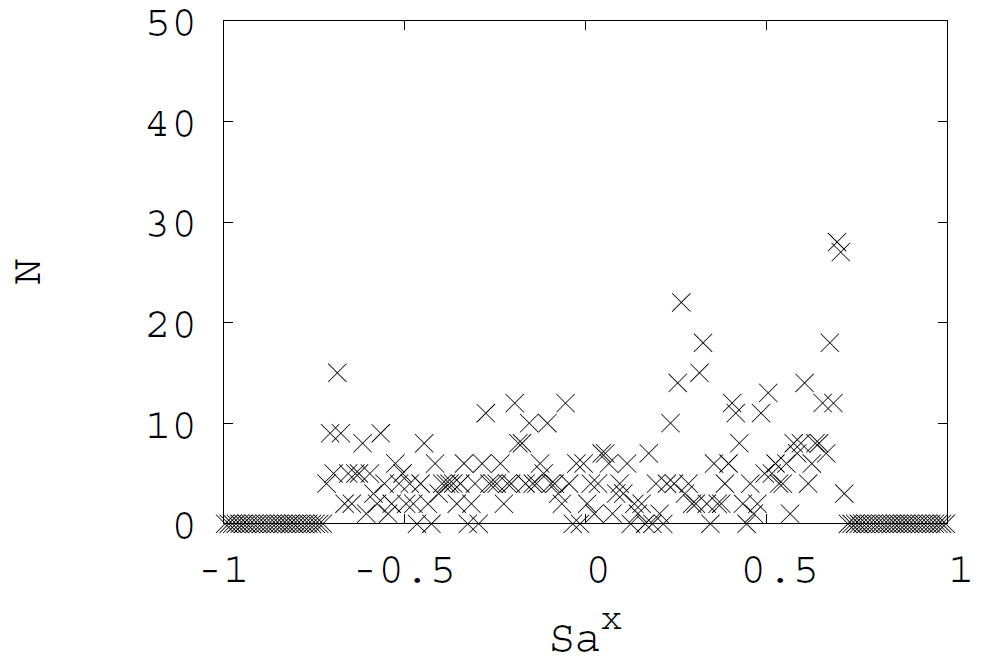}
  \end{center}
 \end{minipage}
\vspace{1.0cm}
\caption{Distributions of $S_a^x$ of two successive kagome layers for $J=0.1$ and $\omega =0.5$.}
\end{figure}
The distribution for one kagome layer seems to be almost the same as that for the nearest neighbor kagome plane after the operation of $S_a^x \rightarrow -S_a^x$.

Similar ordered states appear for $J=0.125, 0.13$ and $0.14$ with $\omega =0.5$. Although the ordered states presented in this subsection are seemingly glassy, more elaborate investigations are necessary to confirm the nature of the orders in the future.

\subsection{Multidomain structure including vortices in hexagons}
Spin structure of the central kagome plane for $J=0.15$ and $\omega =0.5$ is shown in Fig. 9.
\begin{figure}
 \begin{minipage}{0.49\hsize}
  \begin{center}
   \includegraphics[width=65mm]{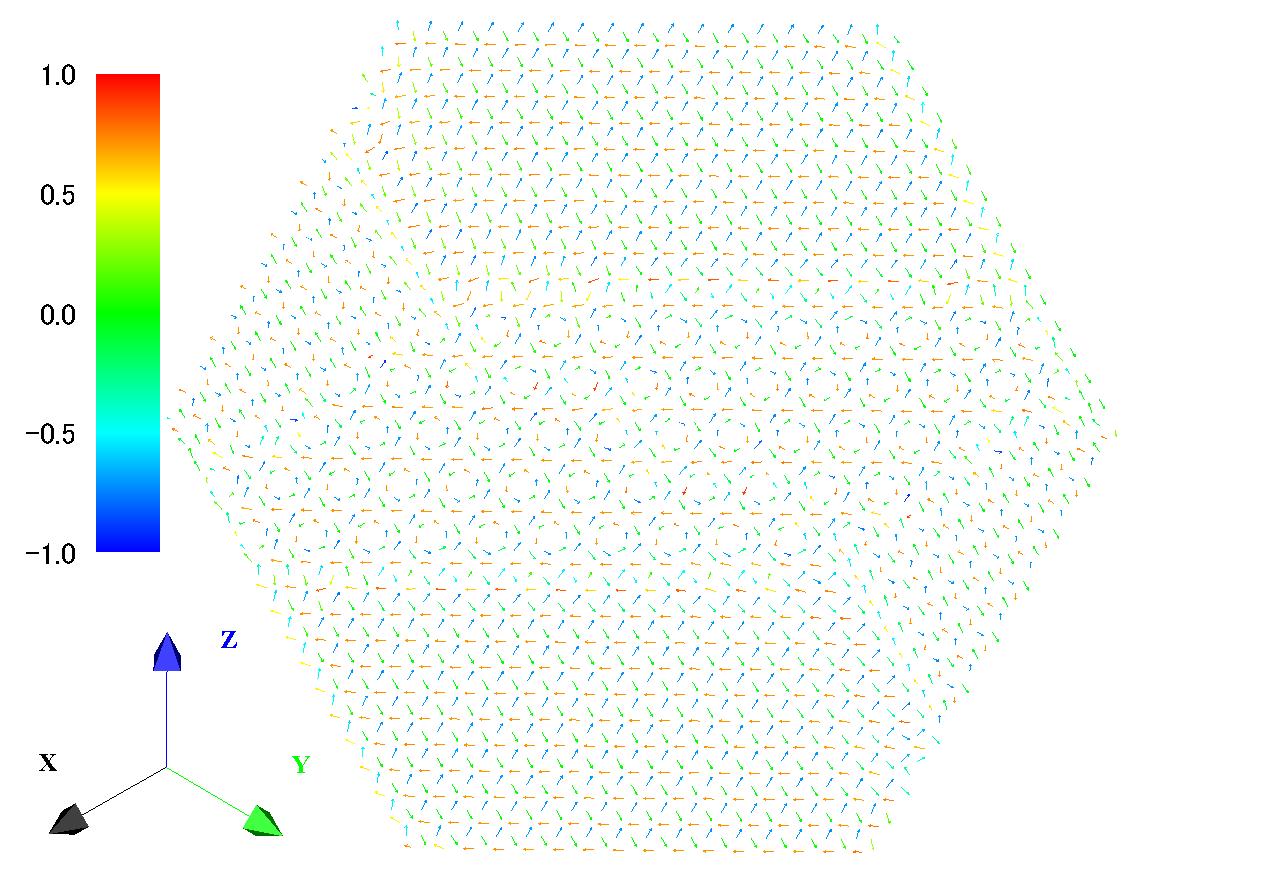}
  \end{center}
 \end{minipage}
 \begin{minipage}{0.49\hsize}
  \begin{center}
   \includegraphics[width=65mm]{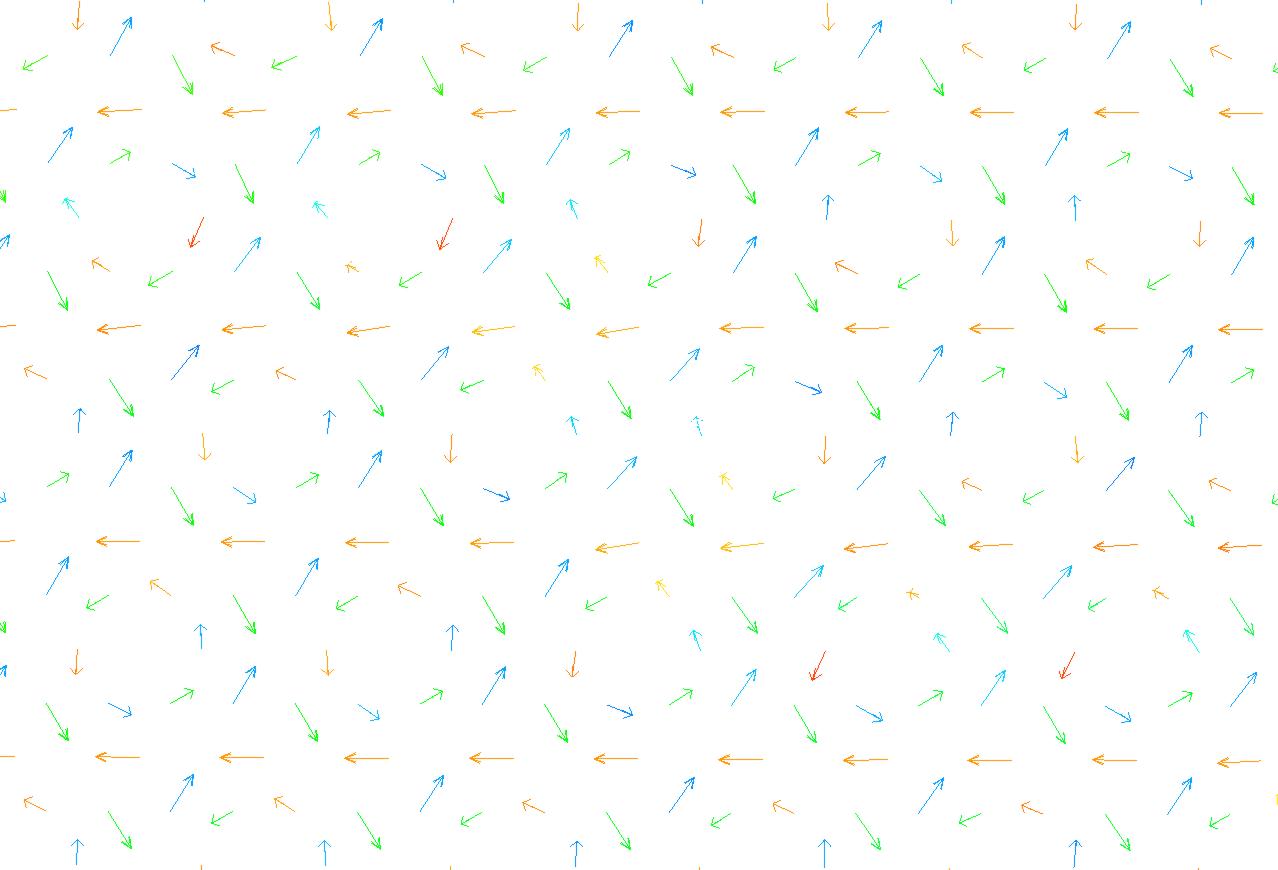}
  \end{center}
 \end{minipage}
\vspace{1.0cm}
\caption{Spin structure of the central kagome plane for $J=0.15$ and $\omega =0.5$. The color legend shows the value of $S^x$. In the enlarged view, clockwise and counterclockwise vortices can be seen clearly.}
\end{figure}
The color legend shows the value of $S^x$. In the central part of the kagome plane, a domain with hexagonal vortices is found. In the enlarged view, clockwise and counterclockwise vortices can be seen clearly. Vortices appear every other hexagons in the kagome plane.

Vorticity vector ${\bf v}$ of a hexagon in a kagome plane may be defined as follows. Six unit direction vectors from the center of a hexagon to the six vertexes are named ${\bf r}_1$ to ${\bf r}_6$. Vorticity vector is
\begin{equation}
{\bf v}\equiv ({\bf r}_1\times{\bf S}_1 +{\bf r}_2\times{\bf S}_2 +\cdots +{\bf r}_6\times{\bf S}_6 )/6 
\end{equation}
where ${\bf S}_i$ is the spin vector of ${\it i}$-th vertex of the hexagon. If ${\bf S}_i$'s are on the kagome plane and vertical to the corresponding ${\bf r}$'s, ${\bf v}$ would be a unit vector perpendicular to the kagome plane. Vorticity may be defined by
\begin{equation}
v={\bf v}\cdot {\bf e}_{111}
\end{equation}
where ${\bf e}_{111}$ is the unit vector of the $[111]$ direction and if the kagome plane is $(111)$. It would be $+1$ for perfect counterclockwise vortex and $-1$ for clockwise one. For the vortexes in the central kagome plane, $\vert v\vert\simeq 0.57\pm 0.02$ and the angles between the clockwise vorticity vectors and $[111]$ are almost $0^\circ$ and those between the counterclockwise vorticity vectors and $[111]$ are almost $180^\circ$. The reason why $\vert v\vert$ is considerably smaller than $1$ is that the spins on the vertexes of the hexagons, which construct vortexes are not on the $(111)$ kagome plane and the angles between the spin vector and  $[111]$ are almost $36^\circ$ and $144^\circ$ alternately for a hexagon. The angles between two nearest neighbor spin vectors on the hexagons are almost $120^\circ$ that is $120.0^\circ \pm 7.3^\circ$. There are hexagons without vortexes between two hexagons with vortexes. For these hexagons, among six angles between two nearest neighbor spins on the same hexagons, four are about $120^\circ$ and other two are about $60^\circ$. Angle distributions of $\theta_{ab}$, $\theta_{bc}$, and $\theta_{ca}$ for the central kagome plane are depicted in Fig. 10 in $2^\circ$ increments.
\begin{figure}
  \begin{center}
   \includegraphics[width=140mm]{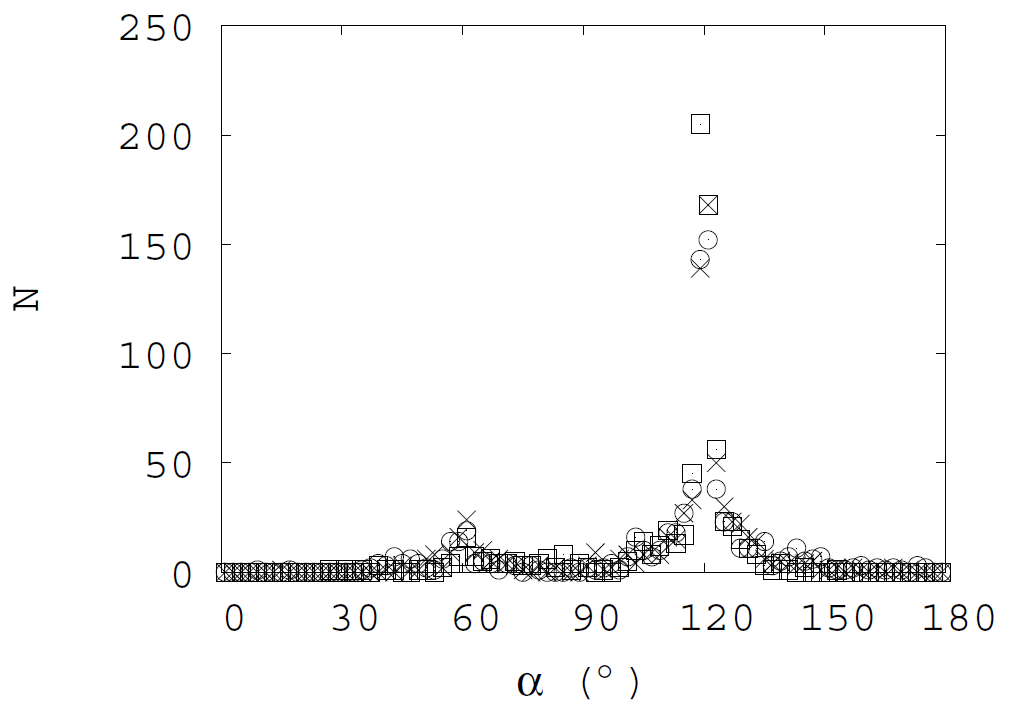}
  \end{center}
\vspace{1.2cm}
\caption{Angle distributions of $\theta_{ab}$, $\theta_{bc}$, and $\theta_{ca}$ for $J=0.15$ and $\omega =0.5$ of the central kagome plane. Crosses, circles, and squares represent angles for $\theta_{ab}$; $\theta_{bc}$, and $\theta_{ca}$, respectively in $2^\circ$ increments.}
\end{figure}
There is a large peak around $120^\circ$ and a small peak around $60^\circ$.

In Fig. 11, distributions of $S_a^x$ for successive five kagome layers are shown.
\begin{figure}
 \begin{minipage}{0.19\hsize}
  \begin{center}
   \includegraphics[width=25mm]{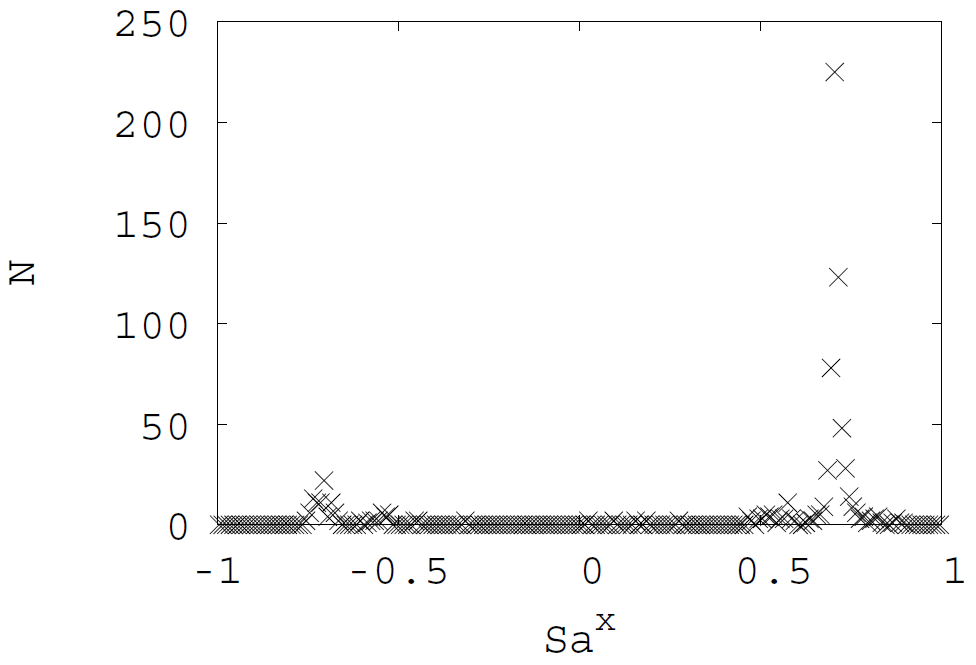}
  \end{center}
 \end{minipage}
 \begin{minipage}{0.19\hsize}
  \begin{center}
   \includegraphics[width=25mm]{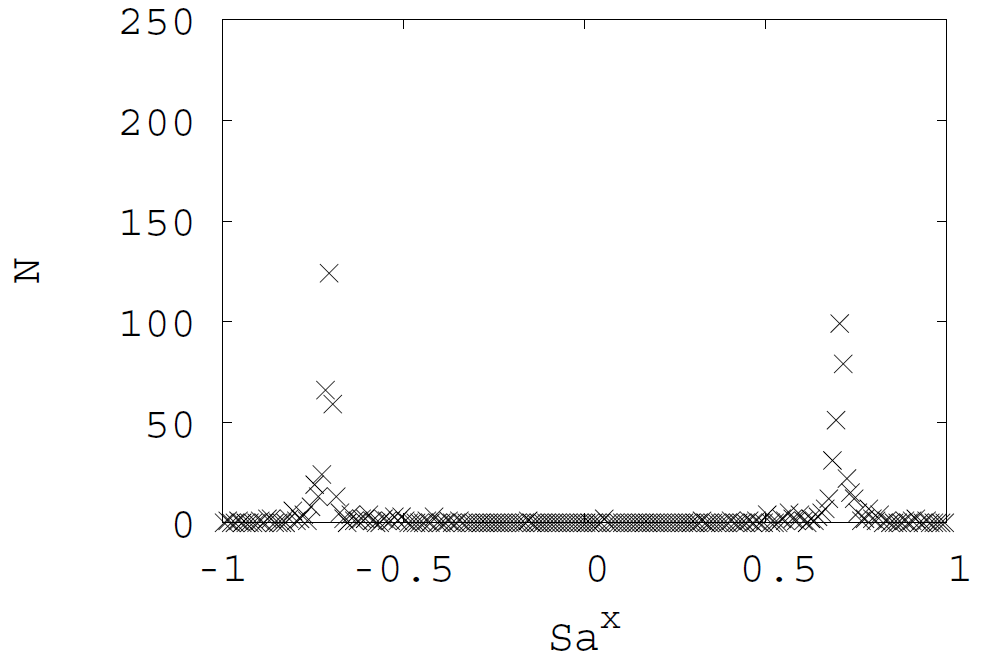}
  \end{center}
 \end{minipage}
 \begin{minipage}{0.19\hsize}
  \begin{center}
   \includegraphics[width=25mm]{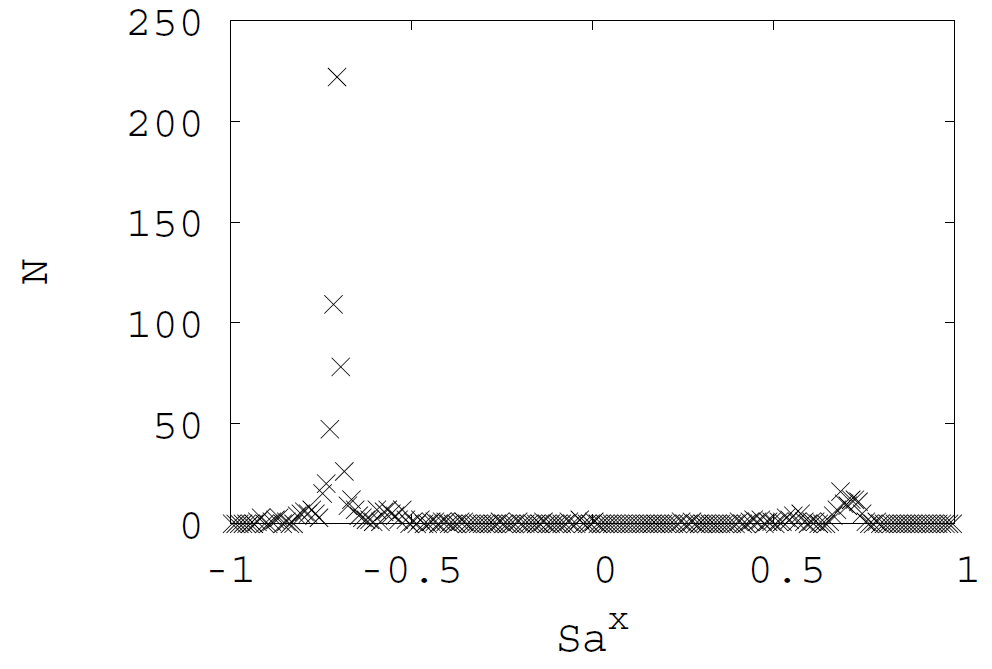}
  \end{center}
 \end{minipage}
 \begin{minipage}{0.19\hsize}
  \begin{center}
   \includegraphics[width=25mm]{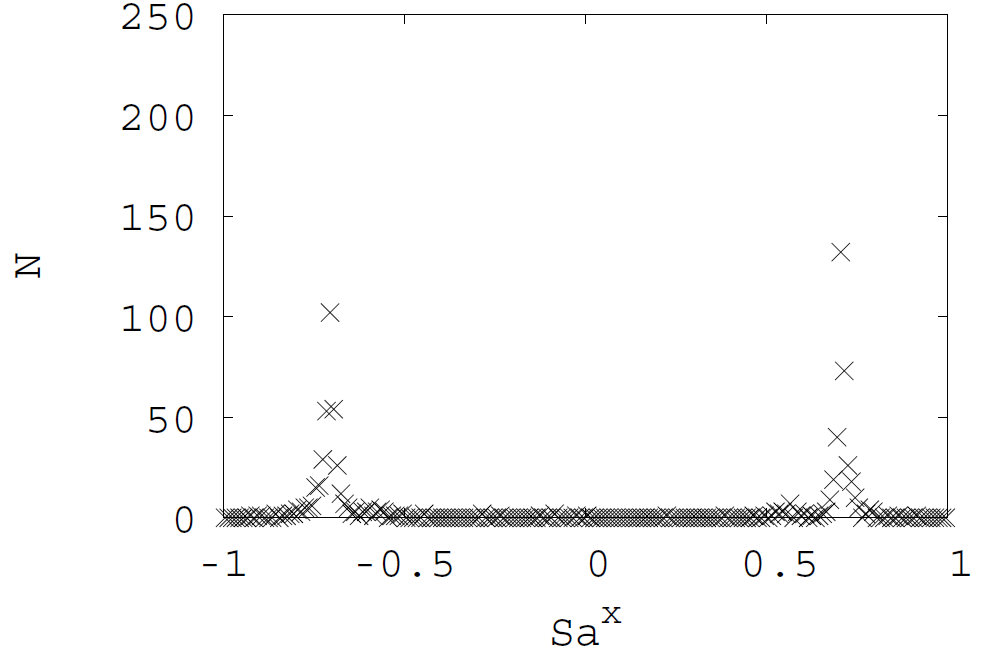}
  \end{center}
 \end{minipage}
 \begin{minipage}{0.19\hsize}
  \begin{center}
   \includegraphics[width=25mm]{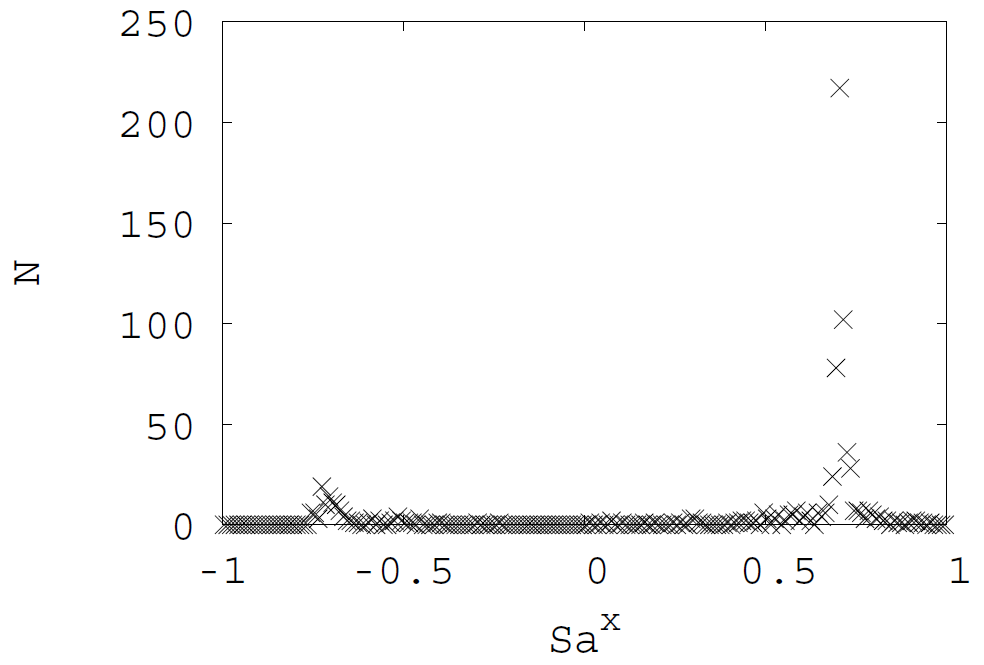}
  \end{center}
 \end{minipage}
\vspace{0.5cm}
\caption{Distributions of $S_a^x$ for $J=0.15$ and $\omega =0.5$ of  successive five kagome layers.}
\end{figure}
 This figure suggests periodicity of nearly $4$ in the $[111]$ direction.

\subsection{Multidomain $120^\circ$ structures with various periodicity along $<111>$ direction}
For systems with larger $J/\omega$, namely from $J/\omega =0.32$ to $0.8$, spin structures seem to be multidomain $120^\circ$ structures. Systems of $J=0.16$, $0.17$, $0.18$, $0.19$, $0.2$, $0.3$, and $0.4$ with $\omega =0.5$ have been investigated. Except for the system with $J=0.18$, multidomain structures in the kagome planes have been observed. Spin structure of the central kagome plane for $J=0.17$ with $\omega =0.5$ is shown in Fig. 12.
\begin{figure}
  \begin{center}
   \includegraphics[width=140mm]{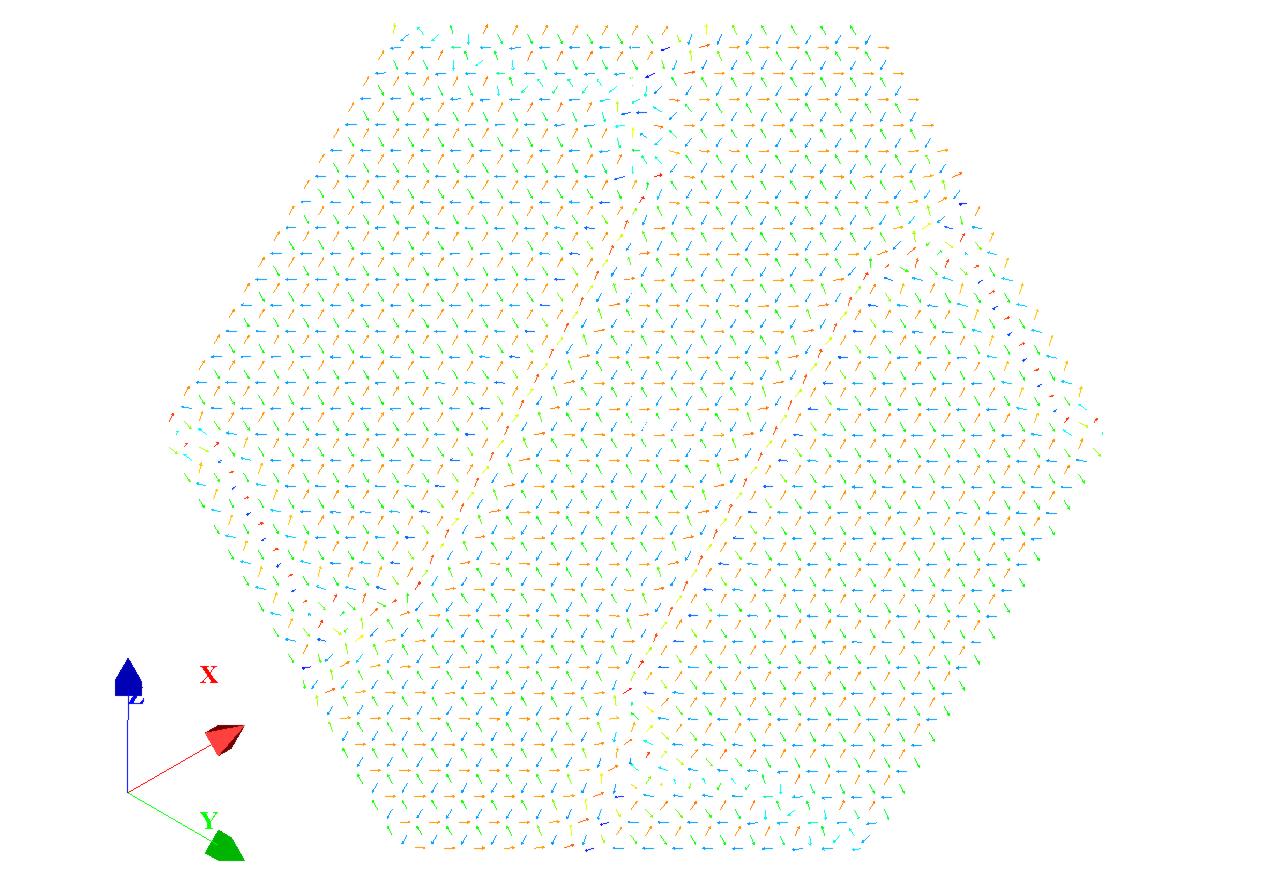}
  \end{center}
\vspace{0.5cm}
\caption{Spin structure of the central kagome plane for $J=0.17$ with $\omega =0.5$. The colors represent the values of $S^x$ in the same way as Fig. 3. }
\end{figure}

The distributions of the angles between the spins have peaks around $120^\circ$ and are shown in Fig. 13 for $J=0.16$ and $0.4$.
\begin{figure}
 \begin{minipage}{0.49\hsize}
  \begin{center}
   \includegraphics[width=65mm]{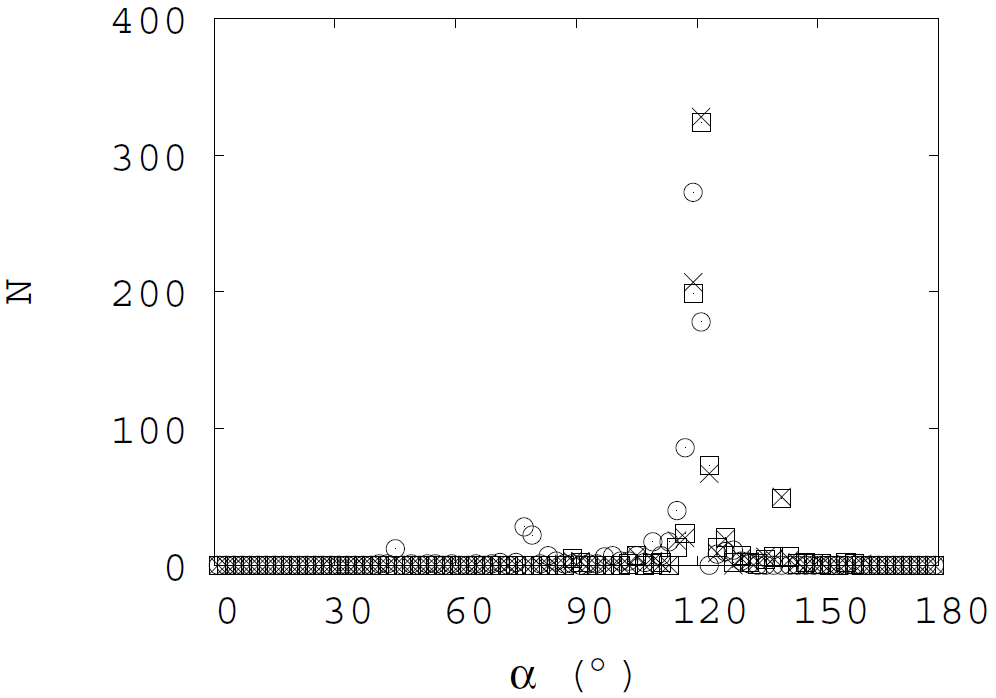}
  \end{center}
 \end{minipage}
 \begin{minipage}{0.49\hsize}
  \begin{center}
   \includegraphics[width=65mm]{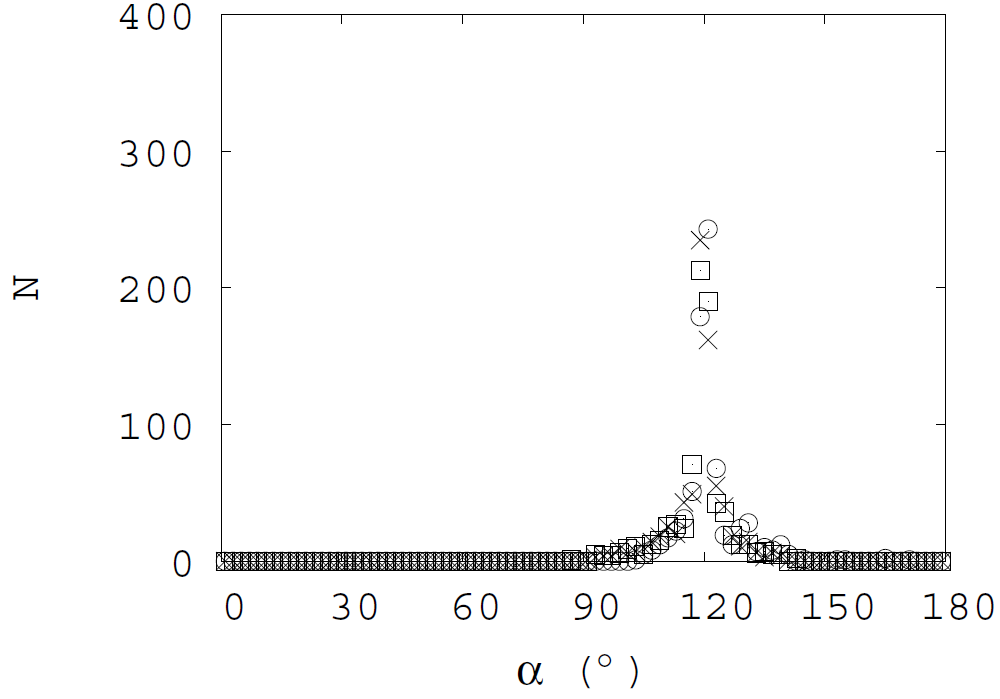}
  \end{center}
 \end{minipage}
\vspace{1.0cm}
\caption{Angle distributions of $\theta_{ab}$, $\theta_{bc}$, and $\theta_{ca}$ for $J=0.16$ (left) and $0.4$ (right) with $\omega =0.5$ of the central kagome plane. Crosses, circles, and squares represent numbers of angles for $\theta_{ab}$; $\theta_{bc}$, and $\theta_{ca}$, respectively in $2^\circ$ increments.}
\end{figure}
This is also the case for systems with other $J$'s. For the system of $J=0.18$ with $\omega =0.5$, the angle distributions are very sharp, which is probably caused by the absence of the domain boundary.

The periodicity of the spin structure may be investigated in the same way as the preceding subsection using the distributions of spin components for successive kagome layers. The periodicity for $J=0.16$ and $0.17$ is about $4$, whereas that for $J=0.18$ is probably between $5$ and $6$ and incommensurate. For $J=0.19$ and $0.2$, it is about $5\pm 2$ and $6\sim 8$, respectively. For $J=0.3$ and $0.4$, it is between $10$ and $11$, and  about $16$, respectively. The periodicity seems to increase as $J/\omega$ increases but it is not conclusive partially because of the existence of domains. 

\section{Discussion and Conclusion}
\begin{table}
\caption{Ordered states for various $J$ and $\omega$. Type of the order is shown.}
\begin{center}
\begin{tabular}{llp{8cm}l}
\hline
$J$ & $\omega$ & type of order &\\
\hline
0.0 & 0.5 & $120^\circ$ order in a $\{111\}$ plane with opposite spin directions in adjacent kagome planes & \\
0.01 & 0.5 & $120^\circ$ order in a $\{111\}$ plane with opposite spin directions in adjacent kagome planes  & \\
0.02 & 0.5 & $120^\circ$ order in a $\{111\}$ plane with opposite spin directions in adjacent kagome planes & \\
0.05 & 0.5 & $120^\circ$ order in a $\{111\}$ plane with opposite spin directions in adjacent kagome planes & \\
0.1 & 0.5 & glassy spin cluster or small multidomain state & \\
0.125 & 0.5 & glassy spin cluster or small multidomain state & \\
0.13 & 0.5 & glassy spin cluster or small multidomain state & \\
0.14 & 0.5 & glassy spin cluster or small multidomain state & \\
0.15 & 0.5 & multidomain including vortices, period of about 4 & \\
0.16 & 0.5 & multidomain $120^\circ$ order with period of about 4 & \\
0.17 & 0.5 & multidomain $120^\circ$ order with period of about 4 & \\
0.18 & 0.5 & $120^\circ$ order with period of $5\sim 6$ & \\
0.19 & 0.5 & multidomain $120^\circ$ order with period of $5\pm 2$ & \\
0.2 & 0.5 & multidomain $120^\circ$ order with period of $6\sim 8$ & \\
0.3 & 0.5 & multidomain $120^\circ$ order with period of $10\sim 11$ & \\
0.4 & 0.5 & multidomain $120^\circ$ order with period of about 16 & \\
0.5 & 0.5 & $120^\circ$ planar order& \\
1.0 & 0.5 & $120^\circ$ planar order& \\
2.0 & 0.5 & $120^\circ$ planar order& \\
2.0 & 0.3 & $120^\circ$ planar order& \\
2.0 & 0.1 & $120^\circ$ planar order& \\
2.0 & 0.01 & $120^\circ$ planar order& \\
2.0 & 0.001 & $120^\circ$ order with defect planes& \\
2.0 & 0.0 & $120^\circ$ planar order with defect planes& \\
\hline
\end{tabular}
\end{center}
\end{table}
Heisenberg antiferromagnets on fcc lattice structure with ABC stacked kagome planes of magnetic ions have been investigated by solving the LL equation numerically. Continuous space approximation has been used to take account of long range nature of the dipole-dipole interactions properly and effectively. Continuous space approximation may be qualitatively adequate from the following examples. Other systems with only the dipolar interactions have been investigated to obtain qualitatively valid ordered states. The ferromagnetic state is obtained in the fcc lattice using the continuous space approximation \cite{y1}, which is consistent with the previous study without using the approximation \cite{bz}. In the pyrochlore lattice, complex ferromagnetic state has been obtained using the continuous space approximation \cite{y2}. Complex ferromagnetic state seems to be obtained without using the approximation \cite{y}. In the pyrochlore lattice, with both exchange and dipolar interactions, the Palmer-Chalker (PC) state \cite{pc} has been obtained in a range of relative strength of the two interactions using the continuous space approximation \cite{y2}, which is consistent with another investigation \cite{mg}. In the PC state, a pair of spins and another pair of spins for each tetrahedron constituting the lattice are antiparallel and parallel to the opposite edge of the tetrahedron.

Besides the layered $120^\circ$ spin structure with defect lines (planes) for the system with only nearest-neighbor antiferromagnetic exchange interaction \cite{tetal,hetal} and the probable $120^\circ$ structure realized by an order-by-disorder process with staggered spin structure in a $<111>$ direction for the system with only dipolar interactions \cite{wetal}, $120^\circ$ spin structures with various periodicity in a $<111>$ direction, spin structure with a vorticity in hexagons of the kagome planes, and possible glassy spin cluster state appear. They are summarized in Table I.

Because of the long range character of the dipolar interactions, it is possible that the system size $L=64$ is not enough to decide precise ordered states especially for small $J/\omega$. Whether the $120^\circ$ staggered spin structure in an order-by-disorder process is exactly realized is a problem that should be further investigated.

For the systems of pure exchange interaction and with very small dipolar interactions, the directions of spins do not seem to be in the kagome plane although they are $120^\circ$ structures. Whether the directions of spins are really not related to the crystal direction is another problem.

Seemingly glassy spin cluster states should be investigated in the future to clarify the dynamical nature of the present frustrated systems.

To search for systems with various spin orders appeared in the present study in magnetic materials would be interesting.

\end{document}